\documentclass[11pt,preprint]{aastex}

\usepackage{graphics,graphicx}
\usepackage{natbib}
\usepackage{epstopdf}
\newcommand{\gtorder}{\mathrel{\raise.3ex\hbox{$>$}\mkern-14mu
            \lower0.6ex\hbox{$\sim$}}}
\newcommand{\ltorder}{\mathrel{\raise.3ex\hbox{$<$}\mkern-14mu
            \lower0.6ex\hbox{$\sim$}}}

\shorttitle{Supermassive black hole binary alignment}
\shortauthors{}

\begin{document}

\title{Alignment of supermassive black hole binary orbits and spins}

\author{M. Coleman Miller\altaffilmark{1}}

\and

\author{Julian H. Krolik\altaffilmark{2}}

\altaffiltext{1}{Department of Astronomy and Joint Space-Science Institute, University of Maryland, College Park MD 20742--2421 USA; Department of Physics and Astronomy, Johns Hopkins University, Baltimore, MD 21218, USA: miller@astro.umd.edu} 

\altaffiltext{2}{Department of Physics and Astronomy, Johns Hopkins University, Baltimore, MD 21218, USA}

\begin{abstract}

Recent studies of accretion onto supermassive black hole binaries suggest that much, perhaps most, of the matter eventually accretes onto one hole or the other.   If so, then for binaries whose inspiral from $\sim 1$~pc to $\sim 10^{-3}-10^{-2}$~pc is driven by interaction with external gas, both the binary orbital axis and the individual black hole spins can be reoriented by angular momentum exchange with this gas.  Here we show that, unless the binary mass ratio is far from unity, the spins of the individual holes align with the binary orbital axis in a time $\sim {\rm few}-100$ times shorter than the binary orbital axis aligns with the angular momentum direction of the incoming circumbinary gas; the spin of the secondary aligns more rapidly than that of the primary by a factor $\sim (m_1/m_2)^{1/2}>1$.  Thus the binary acts as a stabilizing agent, so that for gas-driven systems, the black hole spins are highly likely to be aligned (or counteraligned if retrograde accretion is common) with each other and with the binary orbital axis.  This alignment can significantly reduce the recoil speed resulting from subsequent black hole merger.

\end{abstract}


\section{Introduction}

Supermassive black hole binaries are likely to be formed after the merger of the black holes' host galaxies.  There is considerable discussion, but little conclusive observational evidence, about whether the holes themselves ultimately coalesce \citep{2002Sci...297.1310M, 2003MNRAS.340..411L, 2008Natur.452..851V, 2009Natur.458...53B, 2010ApJ...724L.166I, 2010ApJ...710.1205H, 2010ApJ...717L..37H}.   Whether or not merger is achieved, it does appear to be possible that in gas-rich galaxy mergers dynamical interaction between the binary and surrounding gas could play a key role in shrinking the binary to separations $\sim 10^{-3}-10^{-2}$~pc \citep{2000ApJ...532L..29G, 2002ApJ...567L...9A, 2004ApJ...607..765E, 2005ApJ...630..152E, Kazantzidis05,2009MNRAS.398.1392L,Shi12, 2013MNRAS.429.3114C}; at smaller separations,  gravitational radiation causes the binary to coalesce in less than a Hubble time.  The initial treatments of circumbinary accretion suggested that binary torques could act as a wall to prevent the gas from getting to the individual holes \citep{Pringle91, 1991ApJ...370L..35A, Artym94,Milos05}, but more recent two- and three-dimensional simulations find that $\sim 10$--$100\%$ of the gas is eventually accreted by the holes \citep{1996ApJ...467L..77A, 1997MNRAS.285...33B, 2007PASJ...59..427H, MM08, 2010ApJ...708..485H, Shi12, 2012MNRAS.427.2680K, 2012MNRAS.427.2660K, Noble12, Roedig12, DOrazio13}.  Among other consequences, this means that the orientation of the circumbinary gas can couple to the orientation of the black holes via accretion.

Here we treat quantitatively the alignment torques within the coupled circumbinary disk--binary orbit--black hole spins system.  In Section~2.1 we calculate the rate at which the binary orbit aligns with the axis of a circumbinary disk and the rates at which the individual black hole spins align with the orbital axis through Lense-Thirring torques on their individual accretion disks (i.e., the ``Bardeen-Petterson mechanism": \citealt{BP75}).    Because the formalisms for the two are so similar, we calculate these rates in parallel.   The Lense-Thirring torques tend to be strongest at small radii in the individual disks (which we call ``minidisks"), whereas the tidal torques on each minidisk due to the other black hole are strongest at large radii.    In Section~2.2 we show that in many cases there can be significant overlap between the regions subject to these torques; when this occurs, spin alignment with the orbital axis is accelerated.    In Section~2.3 we show that these alignment rates are usually considerably faster than the rate at which the binary orbital elements evolve due to interaction with surrounding gas.  Thus, for gas-driven systems, the spin axes are likely to be closely aligned (or counteraligned) with the orbital axis at the time of merger.
Only when the binary mass ratio is very far from unity do the spin alignment times become longer than the orbital plane alignment time.
In Section~3 we conclude by discussing the implications of our result for gravitational wave kicks at merger.  Throughout this paper, we assume that dynamical interactions with stars can be neglected.  If instead such interactions are important, significant misalignment is possible because torques from stars can affect the orientation of the binary without changing the orientations of the spins.

\section{Alignment of orbits and spins}

Let the two black holes masses be $m_1$ and $m_2=qm_1\leq m_1$ for total mass $M=m_1+m_2$ and symmetric mass ratio $\eta=m_1m_2/M^2=q/(1+q)^2$.  Note our convention that lower-case $m$ denotes the mass of an individual black hole, whereas upper-case $M$ is the total mass.  They orbit each other with semimajor axis $a$; these and some of our other parameters are defined in Table~1.   At large separations, the eccentricity of the binary could be $\sim 0.6$ or higher \citep{1991ApJ...370L..35A,2003ApJ...585.1024G,2005ApJ...634..921A,2009MNRAS.393.1423C,2011MNRAS.415.3033R,2012JPhCS.363a2035R}, but as we discuss briefly at the end of Section~2.2 such eccentricities make only a moderate quantitative difference, hence we assume circularity for simplicity.

\begin{table}[ht]
\scriptsize
\caption{Definitions for selected quantities}
\centering
\begin{tabular}{cc}
\hline\hline
Symbol & Definition \\ [0.5ex]
\hline
$m_1,\ m_2$ & Masses of the two black holes    \\
$q$ & Mass ratio $q=m_2/m_1$ \\
$M$ & Total mass $M=m_1+m_2$ \\
$\eta$ & Symmetric mass ratio $\eta=m_1m_2/M^2=q/(1+q)^2$ \\
$a$ & Semimajor axis of binary \\
$r_g$ & Gravitational radius of individual black hole, $Gm_1/c^2$ or $Gm_2/c^2$ \\
$R_g$ & Gravitational radius for the binary, $GM/c^2$ \\
$a_*$ & Dimensionless spin parameter of an individual black hole, $a_*=cJ_1/Gm_1^2$ or $cJ_2/Gm_2^2$ where $J_{1,2}$ are the angular momenta \\
$\alpha$ & The ratio of accretion stress to pressure, introduced by \citet{SS73} \\
$\Omega$ & Orbital frequency, $\Omega=(GM/a^3)^{1/2}$ \\
$h$ & Disk half-thickness; $h$ is a function of radius, $h=h(r)$ \\
$f$ & Parameter indicating the efficiency of alignment of gas plane with spin or binary orbital axis \\
$T_{\rm bin}$ & Time required to align the binary orbital plane with the plane of the circumbinary gas \\
$T_{\rm BP}$ & Time required to align a minidisk plane with the orbital plane \\
\hline
\end{tabular}
\end{table}

\subsection{Alignment of orbits and spins by gas torques}

Analyses of the Bardeen-Petterson effect indicate that the warp transition radius, which is the innermost radius at which there is significant inclination to the black hole spin axis and therefore dominates the alignment rate, is approximately where the rate at which aligning angular momentum due to the Lense-Thirring torque is transported outward matches the rate at which misaligned angular momentum from the outer disk is transported inward \citep{NP00,SKH13b}.    Although misaligned angular momentum transport is not well-described by diffusion \citep{Lodato10,SKH13a}, we will nonetheless suppose that, at the order of magnitude level, the rate of this process does scale in the way a diffusive process would, and that the radial scale of misaligned angular momentum gradients near radius $r$ is $\sim r$.   Then the rate at which the local misaligned angular momentum changes due to this mixing is $\sim f\alpha\Omega(h/r)^2$. The factor $f$ encapsulates several uncertainties about the gas alignment process.   One, as we have already noted, is the assumption that this process resembles diffusion with radial gradient scales $\sim r$.   Another is the intrinsic rate of inward misaligned angular momentum flow.   In the original Bardeen-Petterson paper, $f=1$, i.e., radial misaligned angular momentum transport is entirely due to mass accretion.  On the other hand, \cite{PP83} argued that if the stress responsible for accretion acted like an isotropic viscosity, $f \sim \alpha^{-2}$.    In SPH simulations incorporating that assumption, \cite{LP07} and \cite{Lodato10} found that in fact $f \lesssim 3/\alpha$ when, as would be expected, the warp is nonlinear.   Warp nonlinearity is defined by the criterion $|d \hat \ell /d\ln r| < h/r$, for $\hat \ell$ a unit vector in the direction of the angular momentum averaged over a radial shell at $r$.   More recently, on the basis of MHD simulations without any sort of phenomenological viscosity prescriptions, \cite{SKH13b} have argued that $1 \lesssim f \lesssim \alpha^{-1}$.   In their analysis, $f \sim \alpha^{-1} {\cal M}^2$ when it is primarily due to radial mixing motions with no net mass inflow, where ${\cal M}$ is the typical Mach number of radial flows induced by the warp.  \cite{SKH13a} found that warps generically drive transonic radial flows whose Mach number can be either larger or smaller than unity by factors of several.

For an object located at radius $r$ outside an axisymmetric object of radius $R$, planetary dynamics theory (e.g., \citealt{1999ssd..book.....M}) indicates that the precession rate of the line of nodes is
\begin{equation}
\omega_{\rm bin}={3\over 2}J_2\Omega (R/r)^2\; ,
\end{equation}
where the numerical coefficient
\begin{equation}
J_2={1\over{mR^2}}\int_0^R \, \int_{-1}^{+1} \, d(\cos\theta) \, dr \,  \pi r^4 \,  (3\cos^2\theta -1)  \rho(r,\cos\theta)\; .
\label{eq:J2}
\end{equation}
Here $\rho$ is the mass per unit volume.  Note that at a given $r$, $\omega_{\rm bin}$ is independent of the precise choice of the outer boundary $R$ provided it is large enough to contain all the gravitating mass.
The time-averaged mass distribution associated with a circular binary of mass ratio $q$ is two axisymmetric rings; for this configuration, $J_2=-\eta/2$ if one chooses $R=a$.  Thus the line of nodes undergoes retrograde precession, unlike the prograde precession produced by general relativistic frame-dragging.  If the orbit is moderately eccentric there is only a modest change in the coefficient because what drives precession is the time-averaged quadrupole moment of the binary.

At the order of magnitude level, the rate at which aligning angular momentum is delivered is $\sim \omega_{\rm bin} L_\perp$, where $\omega_{\rm bin}$ is the precession rate due to the quadrupole moment and $L_\perp$ is the local misaligned angular momentum \citep{Larwood97}.   However, \cite{SKH13b} have recently shown that in the context of Lense-Thirring torques, the rate of aligning angular momentum delivery can differ from this estimate by a factor of order unity, which we will call ${\cal I}$.   This quantity is composed of two multiplicative factors.  One is a dimensionless integral accounting for the fact that the very strong radial dependence of the precession frequency ($\sim r^{-3}$ for Lense-Thirring precession, $\sim r^{-7/2}$ for classical quadrupolar precession) means that regions of small misalignment at small radius can be disproportionately strong loci of torque; it has the form
\begin{equation}
 \int_0^1 \, dx \, x^{-3/2} \frac{\sin\delta(x)}{\sin\delta_{\rm BP}} \frac{\Sigma(x)}{\Sigma_{\rm BP}}.
\end{equation}
Here the radius has been non-dimensionalized in units of $R_{\rm BP}$, the Bardeen-Petterson alignment radius.    Quantities subscripted ``BP" are evaluated at $R_{\rm BP}$.   These include $\Sigma$, the surface density, and $\delta$, the misalignment angle.  The other factor accounts for the fact that the direction of the angular momentum carried through the disk to the alignment front is not necessarily exactly opposite the direction of misaligned angular momentum; \cite{SKH13b} estimate that this factor is $\simeq 0.5$.   A very similar formalism should apply to the interaction of a binary with its surrounding disk.   The only difference is that the radial coordinate in the dimensionless integral is normalized to $R_{\rm bin}$, the radius out to which the circumbinary disk is aligned with the binary orbital plane, and the power of $x$ in the integrand is $-2$ rather than $-3/2$.

With this refinement, we may estimate
\begin{equation}\label{eqn:binrad}
{R_{\rm bin}\over a}=\left(3 \eta{\cal I}_{\rm bin}\over{4f  \alpha}\right)^{1/2}\left(R_{\rm bin}\over h_{\rm bin}\right)\; .
\end{equation}
Similarly, Lense-Thirring precession produces a transition radius $r_{\rm BP}$ in each minidisk given by 
\begin{equation}\label{eqn:bprad}
{r_{\rm BP}\over r_g}\approx\left(2a_* {\cal I}_{\rm BP} \over{f  \alpha}\right)^{2/3}\left(r_{\rm BP}\over h_{\rm BP}\right)^{4/3}\; .
\end{equation}
We distinguish radii with respect to the center of mass of the system from those within the minidisks by using an upper-case $R$ for the former and a lower-case $r$ for the latter.

The torque on the binary is the radial integral of the precession rate times the misaligned angular momentum i.e.,
\begin{equation}
N_{\rm bin} = -\eta \pi \frac{GMa^2}{R_{\rm bin}} \sin\delta_{\rm bin} {\cal I}_{\rm bin} \Sigma_{\rm bin}.
\end{equation}
For this estimate, we make several simplifying approximations.  We ignore misaligned angular momentum transferred to the binary through accretion, in the expectation that $R_{\rm bin} \gg a$, so that the accreted matter has already been aligned.   We also assume that the orientation of the circumbinary material is nearly constant, as is consistent with recent numerical simulations \citep{2013ApJ...767...37M}.   If it is not (as in the chaotic accretion scenario of  \citealt{2006MNRAS.373L..90K}), any changes will only lengthen the orbital plane alignment time $T_{\rm bin}$.    The Bardeen-Petterson torque is
\begin{equation}
N_{\rm BP} = 4\pi a_* (r_g c)^2 \left(\frac{r_g}{r_{\rm BP}}\right)^{1/2} \sin\delta_{\rm BP} {\cal I}_{\rm BP} \Sigma_{\rm BP}.
\end{equation}

For time-steady disks heated only by local accretion, \cite{SS73} find that when gas pressure exceeds radiation pressure and scattering opacity exceeds free-free opacity (their ``middle region"),
\begin{equation}
\Sigma=2\times 10^7~{\rm g~cm}^{-2}\alpha^{-4/5}({\dot M}/{\dot M}_{\rm Edd})^{3/5}M_8^{1/5}(R/R_g)^{-3/5}.
\end{equation}
Here ${\dot M}_{\rm Edd}=4\pi GM/(\epsilon\kappa c)$ for opacity $\kappa$ and accretion efficiency $\epsilon \equiv L/({\dot M}c^2)$, and $M_8=M/10^8~M_\odot$.    In the outer Shakura-Sunyaev disk region, where gas pressure exceeds radiation pressure and free-free opacity exceeds scattering opacity,
\begin{equation}
\Sigma=9\times 10^7~{\rm g~cm}^{-2}\alpha^{-4/5}({\dot M}/{\dot M}_{\rm Edd})^{7/10}M_8^{1/5}(R/R_g)^{-3/4}.
\end{equation}
In both these estimates we assume $R \gg R_g$.

The similarity in form of these two expressions suggests that we write 
$$\Sigma=\Sigma_0\alpha^{-4/5}({\dot M}/{\dot M}_{\rm Edd})^\gamma M_8^{1/5}(R/R_g)^{-\beta},$$
where $R=R_{\rm bin}$ for the binary torques and $R=r_{\rm BP}$ for the Bardeen-Petterson torques (and for the Bardeen-Petterson torques ${\dot M}\rightarrow {\dot m}$ and $R_g\rightarrow r_g$).  Similarly, $M_8=M/10^8~M_\odot$ for the circumbinary disk but it is $m_1/10^8~M_\odot$ or $m_2/10^8~M_\odot$ for the minidisks.   If, as we shall assume, the transition radius in the circumbinary disk is in the same Shakura-Sunyaev region as the transition radii in the minidisks, then $\Sigma_0$, $\beta$, and $\gamma$ are the same for the circumbinary disk as for the minidisks.   We note that the few simulations that have been performed in which accretion from the inner edge of a circumbinary disk is distributed between the partners in the binary are consistent with ${\dot m}$ being slightly larger for the lower-mass black hole, hence for $q$ significantly smaller than unity it could be that ${\dot m}_2/{\dot m}_{\rm 2,Edd}$ is considerably larger than ${\dot m}_1/{\dot m}_{\rm 1,Edd}$.

Rewriting the torques in terms of this notation, we have
\begin{equation}
N_{\rm bin} = - \pi \eta (ac)^2  \sin\delta_{\rm bin} {\cal I}_{\rm bin} \Sigma_0 \alpha^{-4/5} ({\dot M}/{\dot M}_{\rm Edd})^\gamma M_8^{1/5} (R_{\rm bin}/R_g)^{\beta-1}.
\end{equation}
Similarly, the Bardeen-Petterson torque on a black hole is
\begin{equation}
N_{\rm BP} = 4\pi a_* (r_g c)^2 \sin\delta_{\rm BP} {\cal I}_{\rm BP} \Sigma_0 \alpha^{-4/5} ({\dot m}/{\dot m}_{\rm Edd})^\gamma m_8^{1/5} (r_{\rm BP}/r_g)^{\beta-1/2}.
\end{equation}

The angular momentum of the binary $L_{\rm bin}=\eta M(GMa)^{1/2}$; that of a black hole is $L_{\rm BH}=a_*(G/c)m^2$.  After some manipulation, we find that the characteristic time to align the binary is
\begin{equation}\label{eqn:bintime}
T_{\rm bin} = \frac{1}{\pi} \frac{\alpha^{4/5}}{ ({\dot M}/{\dot M}_{\rm Edd})^\gamma M_8^{1/5}}\frac{T_0}{\sin\delta_{\rm bin}{\cal I}_{\rm bin}} \left(\frac{R_{\rm bin}}{R_g}\right)^{1-\beta} \left(\frac{R_g}{a}\right)^{3/2}.
\end{equation}
where $T_0=c/(G\Sigma_0)=700$~yr for the middle region and 150~yr for the outer region.   Likewise, the Bardeen-Petterson alignment time is
\begin{equation}
T_{\rm BP} = \frac{\alpha^{4/5}}{ 4\pi({\dot m}/{\dot m}_{\rm Edd})^\gamma m_8^{1/5}}\frac{T_0}{\sin\delta_{\rm BP}{\cal I}_{\rm BP}}\left(\frac{r_{\rm BP}}{r_g}\right)^{1/2-\beta}
\end{equation}
The ratio between the timescales is then
\begin{equation}
{T_{\rm BP}\over{T_{\rm bin}}} = \frac{1}{4} \left(\frac{\dot M}{\dot m}\right)^\gamma \left(\frac{m}{M}\right)^{\gamma-1/5} \frac{\sin\delta_{\rm bin}{\cal I}_{\rm bin}}{ \sin\delta_{\rm BP}{\cal I}_{\rm BP}} \left(\frac{r_{\rm BP}}{r_g}\right)^{1/2-\beta} \left(\frac{R_{\rm bin}}{a}\right)^{\beta-1} \left(\frac{a}{R_g}\right)^{\beta+1/2}.
\end{equation}

From our previous expressions we have
\begin{equation}
\left(r_{\rm BP}\over r_g\right)^{1/2-\beta}=\left({2a_* {\cal I}_{\rm BP}\over f  \alpha}\right)^{(1-2\beta)/3}\left({r_{\rm BP}\over h_{\rm BP}}\right)^{2(1-2\beta)/3}
\end{equation}
and
\begin{equation}
\left({R_{\rm bin}\over a}\right)^{\beta-1}=\left({3\eta {\cal I}_{\rm bin} \over 4f \alpha}\right)^{(\beta-1)/2} \left({R_{\rm bin}\over h_{\rm bin}}\right)^{\beta-1}\; .
\end{equation}
In the Shakura-Sunyaev middle region
\begin{equation}
h/r\approx 2\times 10^{-3}\alpha^{-1/10}({\dot M}/{\dot M}_{\rm Edd})^{1/5}M_8^{-1/10}(r/R_g)^{1/20}
\end{equation}
and in the outer region
\begin{equation}
h/r\approx 8.7\times 10^{-4}\alpha^{-1/10}({\dot M}/{\dot M}_{\rm Edd})^{3/20}M_8^{-1/10}(r/R_g)^{1/8}\; .
\end{equation}
The extremely weak dependences on all parameters in a given disk region mean that we can assume $r_{\rm BP}/h_{\rm BP} \sim R_{\rm bin}/h_{\rm bin} \sim 10^3$.  

Applying this result to the spin alignment time, we find that
\begin{equation}
T_{\rm BP} \simeq \frac{10^{2(1-2\beta)}}{4\pi} T_0 \frac{\alpha^{4/5}}{(\dot m/\dot m_{\rm Edd})^\gamma m_8^{1/5} \sin\delta_{\rm BP}} \left(\frac{2a_*}{f\alpha}\right)^{(1-2\beta)/3} {\cal I}_{\rm BP}^{-2\beta/3}.
\end{equation}
For example, in the Shakura-Sunyaev middle region, where $\gamma = -\beta = 3/5$ and $T_0 = 700$~yr, 
\begin{equation}\label{eqn:bptime}
T_{\rm BP} \simeq 1.4 \times 10^6  \frac{\alpha^{4/5}}{(\dot m/\dot m_{\rm Edd})^{3/5} m_8^{1/5} \sin\delta_{\rm BP}} \left(\frac{2a_*}{f\alpha}\right)^{11/15} {\cal I}_{\rm BP}^{2/5}\hbox{~yr}. 
\end{equation}
If typical values of $\dot m/\dot m_{\rm Edd}$ and $\alpha$ are $\sim 0.1$, the spin alignment time is $\sim 1 \times 10^6$~yr.

Similarly, in the same disk zone we find that the binary orbital plane alignment time
\begin{equation}
T_{\rm bin} \simeq 1.4 \times 10^7  \frac{\alpha^{4/5}}{(\dot M/\dot M_{\rm Edd})^{3/5} M_8^{1/5}\sin\delta_{\rm bin}}  \left(\frac{a}{R_g}\right)^{-1/10}  \left(\frac{3\eta}{4f\alpha}\right)^{4/5} {\cal I}_{\rm bin}^{-1/5}\hbox{~yr}.
\end{equation}
In other words, the orbital plane alignment time is nearly independent of the size of the binary in gravitational units, $a/R_g$,
and in this case the dependence on $\alpha$ cancels identically.

Setting all disk aspect ratios to $10^{-3}$ also leads to the result that
\begin{equation}
\left(r_{\rm BP}\over{r_g}\right)^{1/2-\beta}\left(R_{\rm bin}\over a\right)^{\beta-1} = 10^{-(1+\beta)} \frac{ (2a_* {\cal I}_{\rm BP})^{(1-2\beta)/3}}{(\eta {\cal I}_{\rm bin}/2)^{(1-\beta)/2}} (f\alpha)^{(1+\beta)/6}.
\end{equation}
This form leads to 
\begin{equation}
{T_{\rm BP}\over{T_{\rm bin}}} =  \frac{1}{4} 10^{-(1+\beta)} \left(\frac{\dot M}{\dot m}\right)^\gamma  \left(\frac{m}{M}\right)^{\gamma-1/5}  \frac{\sin\delta_{\rm bin}}{ \sin\delta_{\rm BP}}  \frac{{\cal I}_{\rm bin}^{(1+\beta)/2}}{{\cal I}_{\rm BP}^{2(1+\beta)/3}} \frac{(2a_*)^{(1-2\beta)/3}}{(3\eta/4)^{(1-\beta)/2}} (f\alpha)^{(1+\beta)/6} (a/R_g)^{1/2+\beta}.
\end{equation}

In the middle disk zone, the ratio of spin alignment to orbital plane alignment time is then
\begin{equation}
{T_{\rm BP}\over T_{\rm bin}} \simeq 0.2 \frac{(1+q)^{6/5}}{q^{2/5}} \left(\frac{\dot M}{\dot m}\right)^{3/5} \frac{\sin\delta_{\rm bin}}{\sin\delta_{\rm BP}} a_*^{11/15} \frac{{\cal I}_{\rm bin}^{1/5}}{{\cal I}_{\rm BP}^{4/15}} (f\alpha)^{1/15} (a/R_g)^{-1/10},
\label{eq:ratiomid}
\end{equation}
while in the Shakura-Sunyaev outer region the ratio of times is
\begin{equation}
{T_{\rm BP}\over T_{\rm bin}} \simeq 0.3 \frac{(1+q)^{5/4}}{q^{3/8}} \left(\frac{\dot M}{\dot m}\right)^{7/10} \frac{\sin\delta_{\rm bin}}{\sin\delta_{\rm BP}} a_*^{5/6} \frac{{\cal I}_{\rm bin}^{1/8}}{{\cal I}_{\rm BP}^{1/6}}  (f\alpha)^{1/24} (a/R_g)^{-1/4}.
\label{eq:ratioout}
\end{equation}
In both cases, $q$ in these last two expressions should be interpreted as the ratio between the mass of the black hole whose spin is aligning and the mass of the other black hole.  Remarkably, nearly all the parameters in these expressions enter only with very small exponents.   Their only significant dependences are on $a_*$ and $q$.    They also depend on $\dot M/\dot m$, but this ratio is likely always to be $\sim O(1)$.   The reason $T_{\rm BP}/T_{\rm bin}$ depends so weakly on parameters is that quadrupolar and Lense-Thirring torques act in very similar ways: they are both purely precessional torques, and both precession frequencies scale with radius in almost the same fashion.   In the quadrupolar case, $\omega_p \propto r^{-7/2}$, whereas in the relativistic case, $\omega_p \propto r^{-3}$; both are proportional to a single ``strength" parameter ($\eta$ for the quadrupole, $a_*$ for Lense-Thirring).   The only contrast is in the characteristic inner scale of the radial power-law, $a$ for the quadrupole, $r_g$ for Lense-Thirring---and that ratio enters to at most the 1/4 power, and sometimes to only the 1/10 power.

To gain some perspective on this timescale ratio, consider first the situation of near-equal masses, so that $m \simeq M/2$ and $\dot M \sim 2 \dot m$.    In that case, the only remaining parameter with any significant influence on the ratio $T_{\rm BP}/T_{\rm bin}$ is $a_*$.    In the middle region, $T_{\rm BP}/T_{\rm bin} \simeq 0.7a_*^{11/15}(a/R_g)^{-1/10}$, which is $\sim 0.3$ for $a\sim 10^4~R_g$ when $a_* \sim 1$ and smaller when $a_* \ll 1$.   In the outer region, the ratio changes only slightly, to $\simeq a_*^{5/6}(a/R_g)^{-1/4}$, which is $\sim 0.1$ for $a \sim 10^4~R_g$.   Thus, over a wide range of potentially interesting separations ($a \lesssim 10^4R_g$), the black hole spins in an equal-mass binary align with the binary orbital plane roughly an order of magnitude faster than the binary orbital plane aligns with the outer circumbinary disk.

When the mass ratio is far from unity, we need to consider the primary and the secondary black holes separately.   The time to align either spin increases with black hole mass, but relatively slowly, $\propto m^{\gamma-1/5}$.   The secondary's alignment time is therefore shorter than the primary's by a ratio $\sim (m_2/m_1)^s$, with $0.4 \leq s \leq 0.5$, depending on which accretion regime applies.    On the other hand, for fixed accretion rate and total binary mass, the orbital plane alignment time is $\propto \eta^{4/5}$.   In other words, unequal mass ratios permit more rapid orbital alignment because there is less angular momentum whose direction must be changed.   Moreover, because $T_{\rm bin}$'s scaling with $\eta \equiv q/(1+q)^2$ is stronger than $T_{\rm BP}$'s scaling with $q/(1+q)$, the spin alignment time for either black hole, but especially that of the primary, can become longer than the orbital plane alignment time when the mass ratio is extreme.

\subsection{Further spin alignment by binary torques on minidisks}

The analysis in the previous section treated the outer boundary of the minidisks as free.  In reality, the binary exerts torques on the minidisks (e.g., \citealt{1982ApJ...260..780K,1998ApJ...509..819T,Martin09}), and if those torques are sufficient to maintain spin-disk misalignment to smaller radii than the normal Bardeen-Petterson transition radius, the spins will align even faster than we found above.  This effect was also analyzed by \citet{Martin09} with the assumptions that (using our notation) $f=1/\alpha^2$ and the Bardeen-Petterson transition radius is always well inside the radius at which torques from the binary are important.

As a first step, we argue that despite considerable uncertainty in the exact nature of accretion onto black holes in binary black hole systems, it is likely that both minidisks will extend to their tidal truncation radii.  Consider an initial situation in which the holes do not have minidisks.  If the streams from the circumbinary disk have circularization radii at least a few times the radius of the innermost stable circular orbit (ISCO), then in a steady state in which the accretion rate onto the holes equals the rate at which matter is added to the disk, the total angular momentum of the disk (which by assumption has constant mass) increases because the angular momentum added at the circularization radius exceeds the angular momentum drained into the holes from the ISCO.  This increase can only be terminated when the outer part of the disk is far enough away that tidal torque due to the companion can remove angular momentum from the minidisk and transfer it to the binary orbit.

If either minidisk is misaligned with both the orbital plane and its black hole's spin, torques that tend to align the minidisk with the orbital plane compete with Bardeen-Petterson torques that tend to align the minidisk with the black hole spin.  Previous analyses of the forced precession of an annulus of gas by a binary \citep{1982ApJ...260..780K,1998ApJ...509..819T,LO2000} have found that the precession rate of an annulus of gas at radius $r<a$ around the primary due to the torque by the secondary is given by
\begin{equation}
{\omega_p\over{\Omega_{\rm bin}}}=-{3\over 4}\frac{q}{(1+q)^{1/2}} \left(r/a\right)^{3/2} \cos i
\end{equation}
where $i$ is the inclination of the annulus relative to the orbital plane.   If we set this precession rate equal to the Lense-Thirring precession rate 
\begin{equation}
\omega_{\rm prec,L-T}={2a_*\over{(r/r_g)^3}}{c\over r_g}\; ,
\end{equation}
we find that the two are comparable at a radius
\begin{equation}
r_*/r_g \approx \left(\frac{8 a_*}{3q\cos i}\right)^{2/9} (1 + q)^{2/3}(a/R_g)^{2/3},
\end{equation}
where $R_g$ is in terms of $M$.  

The ratio of this characteristic radius to the usual Bardeen-Petterson radius (eqn.~\ref{eqn:bprad}) is
\begin{equation}
r_*/r_{\rm BP} \approx 1 \times 10^{-4} (\cos i)^{-2/9} a_*^{-4/9} (f\alpha/{\cal I}_{\rm BP})^{2/3} (1+q)^{2/3} q^{-2/9}  (a/R_g)^{2/3},
\label{eq:bintorque}
\end{equation}
where have set $h_{\rm BP}/r_{\rm BP} =10^{-3}$, our fiducial value.  If we suppose that $a_* \sim q \sim f\alpha/{\cal I}_{\rm BP} \sim 1$, $r_* < r_{\rm BP}$ so long as $a \lesssim 10^6 R_g$.     This supposition implies that $f$ is as large as it could plausibly be ($\sim \alpha^{-1}$); a smaller value would strengthen this conclusion.    However, the maximum value of $a/R_g$ for which $r_* < r_{\rm BP}$ diminishes if $a_*$ or $q$ are substantially smaller than unity.   A larger value of $h_{\rm BP}/r_{\rm BP}$ would act in the same direction to greater effect.  To evaluate this ratio for the disk around the secondary as it is torqued by the primary, $q$ should be replaced by $1/q^\prime$, where $q^\prime = m_1/m_2$ .  The mass ratio scaling is then $\propto (1+q^\prime)^{2/3} {q^\prime}^{-4/9}$.

\begin{figure}[htb]
\begin{center}
\plotone{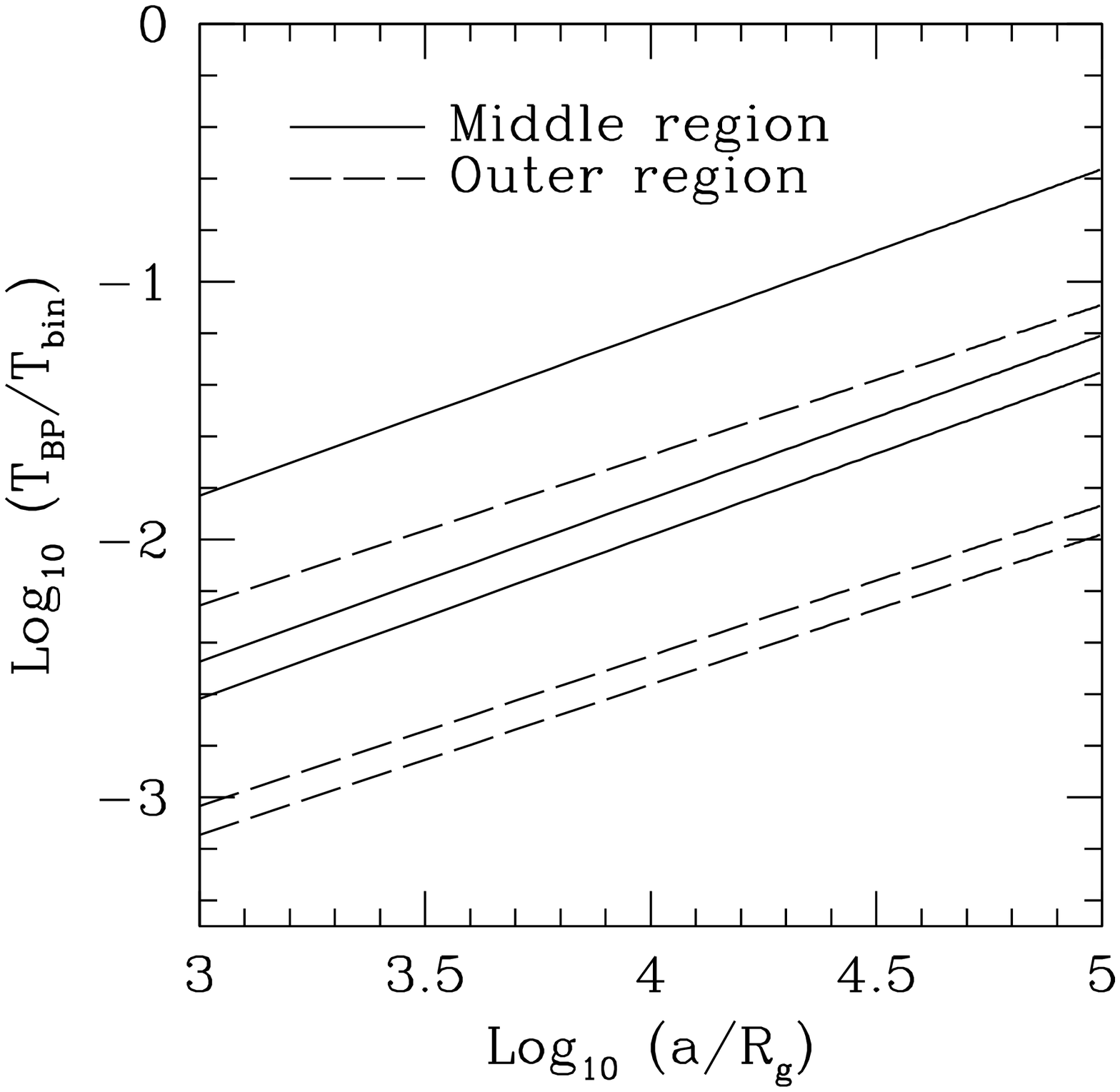}
\caption{Ratio of the spin alignment time to the orbital alignment time, using Equations~(\ref{eq:ratiomid}) and (\ref{eq:ratioout}) along with the correction factor $(r_*/r_{\rm BP})^{1/2-\beta}$ if Equation~(\ref{eq:bintorque}) indicates that $r_*<r_{\rm BP}$.  In this figure the solid lines are for disks in the Shakura-Sunyaev middle region, and the dashed lines are for disks in the outer region.  For both regions, the lowest line (the one with the smallest $T_{\rm BP}$) is for an equal-mass binary with $q=1$, the next lowest has $q=0.1$, and the highest line has $q=10$.  To construct this figure we have assumed that all other factors, e.g., $\sin\delta_{\rm bin}/\sin\delta_{\rm BP}$ in Equations~(\ref{eq:ratiomid}) and (\ref{eq:ratioout}), and $(f\alpha/{\cal I}_{\rm BP})^{2/3}$ in Equation~(\ref{eq:bintorque}), are unity.  From Equation~(\ref{eq:ratioout}), for $T_{\rm BP}$ to be larger than $T_{\rm bin}$ would require that the Eddington ratio for an individual black hole is much smaller than the Eddington ratio for the binary as a whole.  This figure shows that for a wide range of mass ratios and semimajor axes the spins align much faster with the orbit than the orbit does with the circumbinary disk.}
\label{fig:ratio}
\end{center}
\end{figure}

When $r_* < r_{\rm BP}$, binary torques are able to keep the spin and the disk misaligned to smaller radii than would be the case without the torques.    The spins of the individual black holes are then aligned more quickly than we found before because there is material misaligned with respect to the spin at smaller radii than would be found without the binary torques, and the Lense-Thirring effects are much stronger at those small radii.    We previously found that the Bardeen-Petterson torque scales with radius as $R_{\rm T}^{-1/2+\beta}$.  Thus if $\beta=-3/5$ or $\beta=-3/4$, a factor of 10 reduction in $R_{\rm T}$ (corresponding, for example to $a/r_g \sim 3 \times 10^4$) would increase the rate of spin alignment by a factor of 10--20.    At still smaller binary separations relative to $R_g$, the relative speed-up would be even greater.  In Figure~\ref{fig:ratio} we display the final ratio $T_{\rm BP}/T_{\rm bin}$ including extra alignment from binary torques, for three different values of the mass ratio, and for both the middle and outer regions, setting the other factors to unity.  From this figure it is clear that typically the individual black hole spins will be aligned much more rapidly with the orbital axis than the binary orbital axis will be with the circumbinary disk.

We can now address the question of the effect of binary eccentricity on the relative alignment times of the orbit and of the individual black hole spins.  As we discussed after Equation~(\ref{eq:J2}), for the moderate eccentricities $e\sim 0.6$ reached in simulations \citep{1991ApJ...370L..35A,2003ApJ...585.1024G,2005ApJ...634..921A,2009MNRAS.393.1423C,2011MNRAS.415.3033R,2012JPhCS.363a2035R} the torque of the circumbinary disk on the binary orbit will be comparable to what it is for a circular binary.  However, for a given semimajor axis the angular momentum of the binary is less by a factor of $(1-e^2)^{1/2}$ than it is for a circular orbit, hence we expect the binary orbital alignment time to be somewhat shorter.  For $e=0.6$ this factor is 0.8, meaning that the orbital alignment time is not affected much by moderate eccentricities.  

The alignment rate of the individual black hole spins could be increased if the outer radii of the minidisks are smaller than $r_*$ and $r_{\rm BP}$, because then the characteristic radius is reduced and the torque is increased.  This is, however, unlikely to play a significant role for moderate eccentricities.  To see this, note that \citet{2007ApJ...660.1624S} showed that under most circumstances the effective Roche lobe radius for a binary of semimajor axis $a$ and eccentricity $e$ is within 20\% of $(1-e)$ times the standard \citet{1983ApJ...268..368E} value
\begin{equation}
r_{\rm Roche}/a={0.49q^{2/3}\over{0.6q^{2/3}+\ln(1+q^{1/3})}}\; .
\end{equation}
Using our previous expressions, this implies that over the entire range $0<q<\infty$ the ratio of the Roche radius to $r_*$ is within $\sim 30$\% of
\begin{equation}
r_{\rm Roche}/r_*\approx 0.6q^{-4/9}(1-e)\left(a\over R_g\right)^{1/3}\left(3\cos i\over{8a_*}\right)^{2/9}\; .
\end{equation}
As we show in the next section, for $a<{\rm few}\times 10^3~R_g$ gravitational radiation dominates, hence for $q<10$ and $e<0.6$ we expect that $r_{\rm Roche}\gtorder r_*$ and the minidisk alignment rate is not affected much.  Therefore, the eccentricities likely to be reached in these systems have only a small effect on our conclusions.

\subsection{Alignment time vs. binary orbital evolution time}

Even if the minidisks align with the orbital plane much more rapidly than the orbital plane aligns with distant gas, it is still possible that initially misaligned disks could stay misaligned if the binary orbital evolution time is much shorter than the alignment time.  In this section we therefore first demonstrate that when orbital evolution is driven by circumbinary gas, alignment happens on a shorter timescale than the orbit evolves, and hence we expect the black hole spins to be well-aligned with the orbit.  We then explore the opposite extreme, when gravitational radiation dominates orbital evolution, which is relevant at smaller binary separations.

At large separations in our gas-driven scenario, the contraction of the binary as well as its alignment by torques from the circumbinary disk are both driven by the gas.  There have not been sufficient numerical studies to determine how the rate at which the semimajor axis decreases depends on the mass ratio and binary eccentricity, but for a circular equal-mass binary the results of \citet{Shi12} imply 
\begin{equation}
{\dot a\over a}=-0.8{\dot M\over M},
\end{equation}
where as before $M$ is the total mass of the binary.  Thus the binary shrinks on a characteristic time that is comparable to the time needed to increase the mass of the binary.  This time is $\sim 10^8$ years or more for an $e$-folding of mass if accretion proceeds at tens of percent of the Eddington rate.    More quantitatively, if the disks are in the Shakura-Sunyaev middle region,
\begin{equation}
{T_{\rm bin} \over a/{\dot a}} = 0.25 \frac{\alpha^{4/5}}{\sin\delta_{\rm bin}{\cal I}_{\rm bin}^{1/5}} (3\eta/4f)^{4/5} (\dot M/\dot M_{\rm Edd})^{2/5} M_8^{-1/5} (a/R_g)^{-1/10}.
\end{equation}
Because $\eta \leq 0.25$, $\alpha<1$, and $1 \lesssim f \lesssim \alpha^{-1}$, the orbital plane alignment time is at least 1 -- 2 orders of magnitude shorter than the orbital evolution time.  The spin alignment time, even for the primary, is at most comparable to the orbital plane alignment time, so the spin alignment time is even shorter.  Thus, in order for the binary orbital plane and spins to break alignment with the circumbinary disk, the orientation of the circumbinary disk must change on timescales substantially shorter than $M/\dot M \sim a/\dot a$.

If instead the binary is close enough that gravitational radiation is important, then binary shrinkage is decoupled from orbital alignment.
From \citet{1964PhRv..136.1224P}, the characteristic timescale of inspiral of a binary due to gravitational radiation is
\begin{equation}
T_{\rm GW}=a^4(1-e^2)^{7/2}\biggl/\left[{256\over 5}{G^3\eta M^3\over c^5}\right]
\end{equation}
where $e$ is the binary eccentricity.  Numerically, this is
\begin{equation}
T_{\rm GW}=1.3\times 10^6~{\rm yr}(a/10^3 M)^4(0.25/\eta)(M/10^8~M_\odot)(1-e^2)^{7/2},
\end{equation}
where we have scaled to the symmetric mass ratio for equal masses.    Because the time needed to shrink the binary due to gas accretion is $\sim 10^8$~years, independent of $a$,  the semimajor axis at which binary shrinking by gas gives way to shrinking by gravitational radiation is $a\sim {\rm few}\times 10^3 R_g$.

Once the binary begins to evolve more rapidly by gravitational wave emission than by interaction with surrounding gas, we expect the orientation of the binary plane to become fixed.   The reason is that, so long as $R_{\rm bin} > 10^3a$ (cf. eqn.~\ref{eqn:binrad}), $R_{\rm bin}$ will continue to fall well outside the inner edge of the circumbinary disk the entire time until black hole merger.    Our estimate (eqn.~\ref{eqn:bintime}) for the binary orbital alignment time should remain valid.   Because $T_{\rm bin} \propto a^{-3/2}$, the time for the binary orbital plane to respond to changes in the circumbinary disk orientation becomes longer and longer.

At the same time, even if accretion continues, as the results of \citet{Noble12} suggest it will, alignment of the black hole spins also becomes slower.   The effective coupling radius becomes limited by the tidal truncation radius of the disks $r_t$ as the binary becomes tighter.     In addition, if the accretion rate is more than a small fraction of Eddington, when $r_t/r_g \lesssim 10^2$, the disk is radiation dominated, so that its surface density is described by the Shakura-Sunyaev inner region solution,
\begin{equation}
\Sigma \simeq {8/3 \over \alpha} (\dot m/\dot m_{\rm Edd})^{-1} (r/r_g)^{3/2} \kappa^{-1}.
\end{equation}
In these conditions, the nominal black hole spin alignment time becomes
\begin{equation}
T_{\rm BP} = \frac{3}{32\pi} \alpha (\dot m/\dot m_{\rm Edd}) \left[ (r_t/a)^{1/2} (a/r_g)\right]^{-1} \frac{\kappa c/G}{\langle \sin \delta \rangle},
\end{equation}
where $\langle \sin \delta \rangle$ is the inclination angle averaged over radius out to the disk's tidal truncation radius $r_t$.    In a circular binary, the primary's $r_t \simeq 0.4a$ while the secondary's $r_t \sim 0.4qa$ \citep{Artym94}.   Relative to the gravitational wave evolution time,
\begin{equation}
\frac{T_{\rm BP}}{T_{\rm GW}} \simeq 0.1 \alpha (\eta/0.25) (\dot m/\dot m_{\rm Edd}) (r_t/a)^{-1/2} M_8^{-1} \langle \sin \delta \rangle^{-1} (a/10^3 R_g)^{-5}.
\end{equation}
In other words, if the disk were suddenly misaligned from the black hole's spin at about the time that gravitational wave emission begins to dominate the binary's evolution, Bardeen-Petterson spin alignment would be completely ineffective once $a \lesssim 10^3R_g$.   Conversely, whatever mutual relation exists between black hole spins and binary orbit at the time when $a \sim 10^3R_g$ would be preserved until the black holes merge if gas interactions are the only mechanism of reorientation.

\section{Summary and conclusions}

We have shown that then under fairly general circumstances the spins of the individual holes in a supermassive black hole binary will line up with the orbital axis of the binary more rapidly than the orbital axis will line up with the axis of the gas at large distances, and both rates are rapid relative to typical binary evolution times until gravitational wave emission becomes dominant.  The only significant assumption is that the evolution of the binary is driven by interactions with gas rather than with stars.  Hence our conclusions will not apply to gas-poor mergers or situations in which there is a dense cusp of stars around the binary, but should be relevant to a large number of mergers. Thus, even if the gas arrives in small packets with uncorrelated directions, the spin axes are likely to be aligned or counteraligned with each other and with the orbital axis.  Only if the mass ratio is extreme does it become possible for the alignment of black hole spins with the orbital plane to be slower than alignment of the orbital plane with the orientation of the circumbinary disk that is the source of the accretion flow.   In that circumstance, alignment of the primary's spin is also significantly slower than alignment of the secondary's spin.  

If there are circumstances in which the gas arrives in a retrograde direction then there will be a transient phase in which the orbit and spins are turned around, possibly including tearing of the disks; see \citet{2013arXiv1307.0010N} for a recent discussion.  In that case, during this short phase, it could be that there will be substantial misalignment between the various axes.  However, this should occupy only a small fraction of the overall evolution time of the binary and the individual spins, given that completely uncorrelated directions of accretion seem unlikely \citep{2013ApJ...767...37M}.

We note that our conclusion is the opposite of that reached by \citet{2013MNRAS.429L..30L}.  In that paper, the authors did not consider the stabilizing influence of the binary that is the focus of our analysis.  They also effectively decoupled the spin alignment from the binary evolution by assuming that spin alignment depends on the accretion rate whereas the semimajor axis of the binary would shrink on a fixed time (of 10~Myr or 50~Myr).  They then considered a wide range of accretion rates, from $10^{-4}$ to 1 times the Eddington rate with equal realizations in equal logarithmic intervals, such that many of their simulations had little spin alignment but fast shrinkage of the binary.  We therefore feel that our work is a consistent extension of previous arguments that there is likely to be substantial alignment prior to the formation of a gravitationally bound binary \citep{2007ApJ...661L.147B,2010MNRAS.402..682D}.

If there is exact alignment, then upon merger the gravitational wave kick will be less than 200~km~s$^{-1}$ for any mass ratio and spins (see \citealt{2007ApJ...668.1140B,2008ApJ...682L..29B,2010CQGra..27k4006L,2010ApJ...719.1427V,2011CQGra..28k4015Z,2011PhRvL.107w1102L} for fitting formulae for gravitational wave kicks).  These same formulae indicate that even if one black hole is exactly aligned and the other is exactly counteraligned with the orbital axis, the maximum kick is less than about 500~km~s$^{-1}$.  Thus the scenario we have outlined will tend to avoid the ``superkick" configurations of \citet{2011PhRvL.107w1102L}.  This reinforces the suggestion of \citet{2007ApJ...661L.147B} that mergers between gas-poor galaxies are the most likely to lead to kicks high enough to eject the merged supermassive black hole.
 
The question is exactly how aligned a typical system will be upon merger.  Using the kick fitting formulae from \citet{2011PhRvL.107w1102L}, we find that when both spin parameters are drawn uniformly from (0,0.95), the mass ratio is drawn uniformly from (0.25,1), the angles to the orbital axis are drawn uniformly from 0 to 10 degrees, and the azimuthal angles are uniform from 0 to 360 deg, 3\% of kicks exceed 500 km~s$^{-1}$.  When the angles relative to the orbital axis are drawn from 0 to 5 degrees, the fraction drops to 0.024\%.  When the angles are drawn from 0 to 2 degrees, the fraction drops to 0, and the fraction above 200 km~s$^{-1}$ is only 3.6\%.  If one or both spins are retrograde to the orbital axis the fraction of higher kicks increases, but still the kicks are less than 500~km~s$^{-1}$ if the spin axes are within 2 degrees of the orbital axis.  If the primary is mostly aligned with the orbital axis and the mass ratio is roughly in the range of 0.5 to 0.9, then post-Newtonian spin-orbit coupling will align the spin axes further \citep{2004PhRvD..70l4020S,2010ApJ...715.1006K}.  Thus, in concert with other alignment mechanisms previously proposed (e.g., \citealt{2007ApJ...661L.147B}), it is likely that supermassive binary black holes whose orbital evolution is driven by gas torques will have spin axes aligned fairly closely with their orbital axis, leading to suppressed recoil upon merger.

\acknowledgements

This work was supported in part by a grant from the Simons Foundation (grant number 230349 to MCM) and by National Science Foundation grant AST-1028111 (JHK).  MCM also thanks the Department of Physics and Astronomy at Johns Hopkins University for their hospitality during his sabbatical.  We appreciate helpful comments from Laura Blecha, Tamara Bogdanovi\'c, and our referee Umberto Maio.

\bibliography{align}

\end{document}